\documentclass{jfm}

\usepackage{graphicx}
\usepackage{newtxtext}
\usepackage{newtxmath}
\usepackage{natbib}
\usepackage{hyperref}
\hypersetup{
    colorlinks = true,
    urlcolor   = blue,
    citecolor  = black,
}

\newcommand{\RomanNumeralCaps}[1]
\linenumbers

\usepackage{soul}

\usepackage{xcolor}

\title{Crystal growth at a liquid-liquid interface upon drop impact}

\author{Marion Berry\aff{1}  \and Christophe Josserand\aff{2}  \and Anniina Salonen\aff{3}  \and François Boulogne\aff{1}
  \corresp{\email{francois.boulogne@cnrs.fr}}}

\affiliation{\aff{1}Laboratoire de Physique des Solides, UMR 8502, CNRS, Université Paris-Saclay, 91405 Orsay, France.
\aff{2}Laboratoire d’Hydrodynamique (LadHyX), UMR 7646 CNRS-Ecole Polytechnique, IP Paris, 91128 Palaiseau, France.
\aff{3}Soft Matter Sciences and Engineering, ESPCI Paris, PSL University, CNRS, Sorbonne Université, 75005 Paris, France}

\begin{document}
\maketitle

% MAX 250 words
\begin{abstract}
The crystallisation that occurs when a drop is placed in contact with a cold surface is a particularly challenging phenomenon to capture experimentally and describe theoretically.
The situation of a liquid-liquid interface, where crystals appear on a mobile interface is scarcely studied although it provides a defect-free interface.
In this paper, we quantify the dynamics of crystals appearing upon the impact of a drop on a cool liquid bath.
We rationalize our observations with a model considering that crystals appear at a constant rate depending on the thermal shock on the expanding interface.
This model provides dimensionless curves on the number and the surface area of crystals that we compare to our experimental measurements.
\end{abstract}

\begin{keywords}
%Authors should not enter keywords on the manuscript, as these must be chosen by the author during the online submission process and will then be added during the typesetting process (see \href{https://www.cambridge.org/core/journals/journal-of-fluid-mechanics/information/list-of-keywords}{Keyword PDF} for the full list).  Other classifications will be added at the same time.
\end{keywords}

%{\bf MSC Codes }  {\it(Optional)} Please enter your MSC Codes here

%%%%%%%%%%%%%%%%%%%%%%%%%%%%
%
%%%%%%%%%%%%%%%%%%%%%%%%%%%%
\section{Introduction}

When a liquid drop is placed in contact with a material at a temperature lower than the drop melting temperature, crystals grow at the interface until an ice layer is formed, and then, a solidification front propagates away from the interface \citep{Schremb2017,Thievenaz2019}.
On solids, the apparition of crystals has been found to have a key role on drop spreading dynamics.
When crystal growth is faster than the contact line velocity, \citet{Herbaut2019} reported a stick-slip behaviour of the contact line when its velocity is imposed and \citet{Ruiter2017} even observed the arrest of the spreading.
To predict contact line arrest, understanding the development of crystals appears to be crucial.
To this end, recent efforts have been made to visualise the fast crystal growth with techniques involving total internal reflection imaging \citep{Kant2020,Koldeweij2021} and polarised light imaging \citep{Grivet2022}.
Their observations are supported by models based on classic nucleation theory.

Crystals can also appear at a liquid-liquid interface, which provides a soft, defect-free, and stress-free interface \citep{Elsen2013}, in contrast with solid surfaces where surface defects and nanobubbles can be sites for crystal nucleation \citep{Schremb2017a}.
The liquid-liquid configuration is scarcely studied although it is present in applications such as the synthesis of nanocrystals  \citep{Rao2008}, the solidification of emulsion droplets \citep{Denkov2015,Guttman2016}, or the production of particles \citep{Chan2009,lee2015}.

In the present study, we consider thus a drop impacting a thin liquid film on which the crystals forming at the liquid-liquid interface can be visualised.
We have shown in a former work that the growth velocity of these crystals is responsible for the final morphology of impacted drops on the liquid bath \citep{Berry2024}.
Here, we choose to focus on thin liquid films, where the drop spreads radially rather than as an hemispherical cavity formed upon impact on thicker baths.

The article is organized as follows.
In Section~\ref{sec:mat_meth_obs}, we describe the experimental setup and the observations.
Then, we present a model to predict the dynamics of the number of crystals and their surface area in Section~\ref{sec:model}.
Finally, we discuss the results in Section~\ref{sec:discussion} and we conclude.

%\newpage\clearpage
%%%%%%%%%%%%%%%%%%%%%%%%%%%%
%
%%%%%%%%%%%%%%%%%%%%%%%%%%%%

\section{Experimental procedure and observations}\label{sec:mat_meth_obs}

%%%%%%%%%%%%%%%%%%%%%%%%%%%%
\subsection{Materials and methods}

The experiment consists of releasing a drop at room temperature, $T_{\rm d}$, onto a liquid film at a temperature $T_{\rm f}$ lower than the melting temperature $T_{\rm m}$ of the drop.
We choose two non-miscible liquids.
The drop is hexadecane with a molar mass $M_{\rm d}=$ 226.45 g/mol, a melting temperature $T_{\rm m}$ = 18.1~$^\circ$C, a liquid density $\rho_{\rm d}=$ 743 kg/m$^{3}$, a viscosity $\mu_{\rm d}=$ 3.8 mPa.s, a surface tension $\gamma_{\rm d}=$ 27 mN/m, a specific heat capacity $C_{p, \rm d} = 2.20$ kJ/kg K, a thermal conductivity $\lambda_{\rm d} =$ 0.14 W/mK, and an enthalpy of solidification $\mathcal{L}=$ 236 kJ/kg.
We also introduce the surface tension between the liquid phase and solid phase of hexadecane, $\sigma_{ls} = 0.0068$ J.m$^{-2}$ \citep{oliver1975}.
The liquid and thermal properties of hexadecane are taken at room temperature, $T_{\rm d} = 20~^\circ$C \citep{Berry2024}.
The liquid film is a brine constituted of  23.3 wt \% NaCl in pure water.
The film temperature $T_{\rm f}$ is measured with a K-type thermocouple (Radiospare) and ranges from room temperature, $20~^\circ$C to $-21~^\circ$C.
The liquid and thermal properties of the brine are taken at a mean temperature of $0~^\circ$C, giving a density $\rho_{\rm f}=$ 1184 kg/m$^3$, a viscosity $\mu_{\rm f}=$ 2.6 mPa.s, a specific heat capacity $C_{p, \rm f} = 3.3$ kJ/kg K, and a thermal conductivity $\lambda_{\rm f} =$ 0.55 W/mK.
All the liquid properties are summarized in Appendix \ref{appA} and the variation of the properties with temperature is discussed in the Supplementary Materials of \citep{Berry2024}.

The drops are produced with a syringe pump, out of an 18G needle, which provides a drop diameter of $2r_0 = 2.70 \pm 0.04$~mm, measured by image analysis.
A drop is released at an initial height $h_0$ ranging from 5 cm to 50 cm, and the impact velocity is calculated as $v_0 = \sqrt{2g h_0}$ where $g$ is standard gravity.

The solid surface below the liquid film is a 4 mm thick glass slide on top of a silicon wafer.
The glass slide has similar thermal properties as the salt solution and is therefore used to ensure some thermal continuity.
The optical reflectivity of the wafer enhances the visualisation.
The wafer and the glass slide are placed in an aluminium Petri Dish and covered with the salt solution.
We measured the film thickness with an optical device (Chromapoint, STIL).
We choose to work with a film thickness of $h_{\rm f} = 0.27 \pm 0.02$~mm, which gives a dimensionless liquid thickness $H = h_{\rm f} / 2 r_0 = 0.1$.
For such dimensionless thickness, the bath is defined as a liquid film in the literature (\cite{Cossali1997,Tropea1999,Wal2006,Motzkus2009}).

The cooling system is composed of a Peltier module (Adapative, RS) rated at 340 W.
The heat exchange between the Peltier module and the Petri dish is ensured by a home-made receptacle in copper covering the bottom and the perimeter of the Petri dish.
On the other side of the Peltier module, heat is extracted by a heat sink (Laird Technologies) in which a refrigerant liquid circulates at a temperature of $1~^\circ$C by a refrigerated bath (VWR, AP15R-40).

The scene is illuminated with a LED panel (HSC Backlight, Phlox) also positioned at $45^\circ$ pointing toward the liquid film in front of a high-speed camera.
The high-speed camera (FASTCAM Nova S9 1024$\times$1024px, Photron) records at 6000 fps from a 3/4 point of view to measure the projected radius over time $r(t)$, to count the number of crystals $N_{\rm c}$ and to estimate the time $t_{\rm layer}$ to form a frozen layer.
We obtain a resolution of 48 pixels/mm, which is the limiting factor in our experiments.
Nevertheless, as we will show in the next section, this resolution is sufficient to identify the crystals, typically 1 or 2 ms after the impact.

The temperature at the interface $T_{\rm c}$ is constant in the case of two semi-infinite media and is given by \citep{Boeker2011}
\begin{equation}
    T_{\rm c} = \frac{T_{\rm d} + T_{\rm f} e_{\rm f}/e_{\rm d}}{1+e_{\rm f}/e_{\rm d}},
    \label{eq:Tcontact}
\end{equation}
where we used the effusivity $e_i = \sqrt{\lambda_i \rho_i C_{p,i}} =  \lambda_i/\sqrt{D_i}$.
The hypothesis of a semi-infinite phase for the drop is correct for timescales smaller than the heat diffusion timescale, which is of the order of magnitude of a second for our typical length scale, whereas the timescale of impact is tens of milliseconds.
Therefore, we will characterise the thermal shock by $\Delta T_{\rm mc} =  T_{\rm m} - T_{\rm c}$  \citep{Berry2024}.

%%%%%%%%%%%%%%%%%%%%%%%%%%%%
\subsection{Observations}\label{sec:obs}

\begin{figure}
    \centerline{\includegraphics[width=\linewidth]{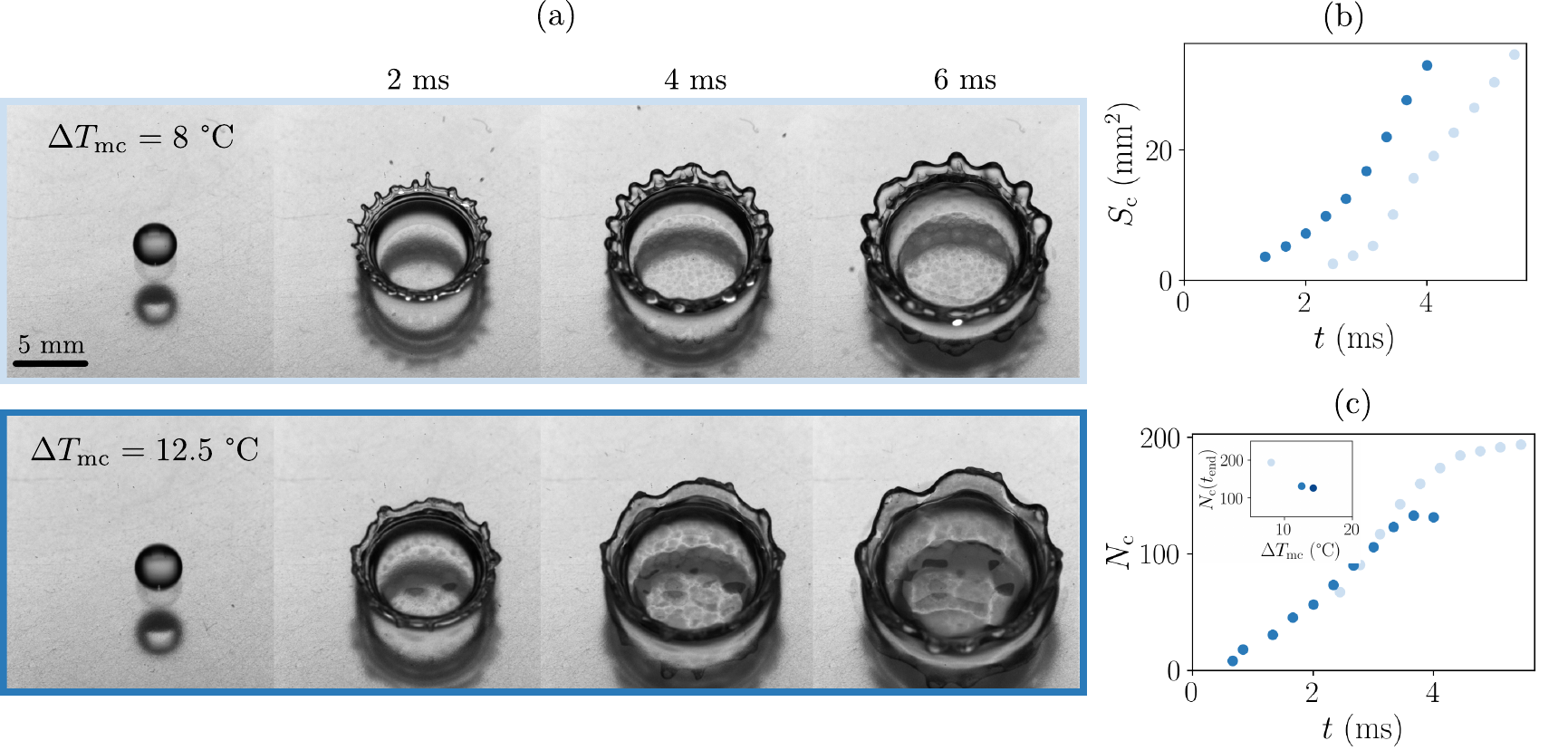}}
    \caption{(a) Time series showing a hexadecane drop impinging a liquid bath at a velocity $v_0$ = 2.8 m/s for two different thermal shocks $\Delta T_{\rm mc} =\{8, 12.5\}~^\circ$C.
    Time evolution of the experimental measurements of (b) the area covered by crystals and (c) the number of crystals, associated to the time series presented in (a) with light blue for $\Delta T_{\rm mc} =8~^\circ$C and dark blue for $\Delta T_{\rm mc} =12.5~^\circ$C.
    The associated movies are provided in Supplementary material with an overlay of the crystals' outlines used for measurements.
    The inset of (c) shows the final number of crystals for different thermal shocks at $v_0 $ = 2.8 m/s.
    }
    \label{fig:observations}
\end{figure}

Figure \ref{fig:observations}(a) shows typical experiments at a fixed impact velocity $v_0$ = 2.8 m/s for two different thermal shocks.
Upon impact, the drop forms a liquid crown analogous to Edgerton's observations \citep{Edgerton1954}.
Additionally, we notice grey clusters that we attribute to crystals appearing at the drop-film interface \citep{Berry2024}.
In time, these crystals grow and additional crystals appear as the drop expands radially.
The time series also show that for a cooler film, \textit{i.e.} a larger thermal shock $\Delta T_{\rm mc}$, fewer crystals are observed, which is counter-intuitive since lower temperatures should enhance crystal nucleation!

To verify this observation, we repeated these experiments at different thermal shocks $\Delta T_{\rm mc} = \{8, 11.2, 12.5, 14.5\}~^\circ$C, for which we measured the surface area of the crystals and counted the number of crystals.
The number of crystals and their area are measured by drawing manually their outline on the images with the software ImageJ.
We cannot obtain measurements before 0.5 ms due to the limited resolution of the camera, and curvature of the drop that causes shadows and optically enlarges the surfaces.
Additionally, near the end of the coverage, it becomes difficult to identify individual crystals because they are very close to one another.
Therefore, as the boundary between crystals is not clear enough, we stop measuring the covered surface before the number of crystals.

These measurements are reported in figures \ref{fig:observations}(b) and (c), where we plot, for two different thermal shocks, the time evolution of the surface area covered by the crystals and the number of crystals, respectively.
For the sake of clarity, error bars are not plotted on these curves but we note that uncertainty for $S_{\rm c}$ and $N_{\rm c}$ would be of 10 \% on each data point.
We observe that increasing the thermal shock (darker blue points) leads to faster dynamics.
In the inset of figure \ref{fig:observations}(c), more measurements of the final number of crystals are given, highlighting a decrease with thermal shock.
For larger thermal shocks, crystals would appear earlier and grow faster, leading to the observation of fewer crystals.
The purpose of the next Section is to establish a model able to render these experimental findings and particularly to explain more quantitatively why the cooler the liquid, the fewer the number of crystals nucleated.

%\newpage\clearpage
%%%%%%%%%%%%%%%%%%%%%%%%%%%%
%
%%%%%%%%%%%%%%%%%%%%%%%%%%%%
\section{Model}\label{sec:model}

\subsection{Dynamics of contact opening}\label{sec:hydro}

In this section, we describe the impact of a drop on the liquid film in terms of contact surface area and timescale to reach maximum opening.
Classically, drop impact on thin films can be characterized by a combination of Reynolds and Weber numbers \citep{Cossali1997,weiss_single_1999,wal_splash_2006,Roisman2008,chen2017}.
The Reynolds number is defined as ${\rm Re} = 2{\rho_{\rm d} v_0 r_0}/{\mu_{\rm d}} $, with $\mu_{\rm d}$ the viscosity of the drop, and varies between 936 and 1872 in this study.
The Weber number is defined as ${\rm We} =  2 \rho_{\rm d} v_0^2 r_0 / \gamma_{\rm d}$, with $\gamma_{\rm d}$ the surface tension of the drop, and varies between 145 and 578 in this study.

\begin{figure}
    \centering
    \includegraphics[width=1\linewidth]{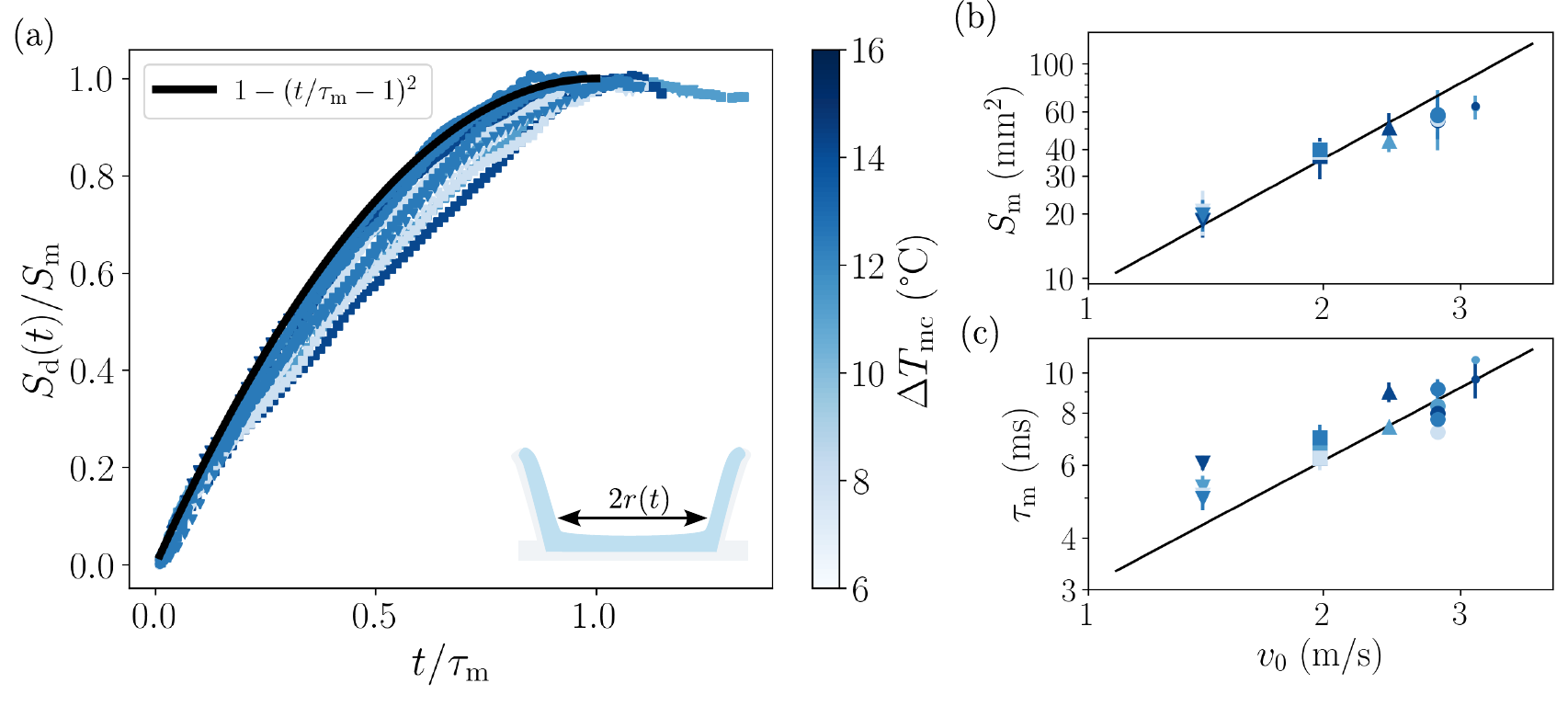}
    \caption{(a) Time evolution of the drop contact area $S_{\rm d}(t) = \pi r(t)^2$ upon impact on a liquid film, for different substrate temperatures and impact velocities.
    Each scatter curve represents one experiment and the black line represents equation \eqref{eq:S_drop}.
     (b) Maximum contact area between the drop and the film and
     (c) time to reach the maximum surface as a function of the impact velocity, with each point representing an experiment. The different markers correspond to impact velocity.
     The descriptions of \cite{Roisman2008} adapted for thin films are plotted by solid black lines.
    }
    \label{fig:hydro}
\end{figure}

We measured through image analysis the dynamics of the surface area $S_{\rm d}(t)$ that corresponds to the contact between the drop and the liquid film as illustrated in figure~\ref{fig:hydro}(a).
During the opening phase of contact, $S_{\rm d}(t)$ grows linearly at short times, which agrees with the literature \citep{weiss_single_1999,Roisman2008} and quadratically to saturation, as shown in figure~\ref{fig:hydro}(a).
For a more accurate description of the whole dynamic until the extension reaches the maximum surface area $ S_{\rm m}$ at a corresponding impact time of $\tau_{\rm m}$, we choose to write the surface area of contact as

\begin{equation}
    \frac{S_{\rm d}(t)}{S_{\rm m}} = 1 - \left( \frac{t}{\tau_{\rm m}} - 1\right)^2,
    \label{eq:S_drop}
\end{equation}
\noindent which is valid for $t < \tau_{\rm m}$.

\cite{Roisman2008} proposed a description for the crown radius on miscible liquid films $H \in$ [0.5, 2], which is also used on immiscible liquid films $H \in$ [0.1, 0.5] by \cite{bernard2021}.
For simplicity, we choose to adapt their expressions for  $H \rightarrow$ 0, which gives dimensionless expressions as
\begin{subequations}
\begin{align}
     {S_{\rm m}} / (\pi r_0^2) &\sim  \, (\beta\sqrt{H}/2)^2 {\rm We},\label{eq:Smax}\\
      \tau_{\rm m} \frac{v_0}{2 r_0}   &\sim \, (\beta{H}/6) {\rm We},\label{eq:tmax}
\end{align}\label{eq:Smax_tmax}
\end{subequations}
where $\beta \sim H^{-1/3}$.  Since we only vary the impact velocity, we have ${S_{\rm m}} \propto v_0^2$ and  ${\tau_{\rm m}}\propto v_0$, which is compared to our data in figure~\ref{fig:hydro}(b,c).
This description is acceptable although the slope against velocity is slightly overestimated, which is not caused by the approximation $H\rightarrow 0$.

%%%%%%%%%%%%%%%%%%%%%%%%%%%%
\subsection{Crystal growth velocity}\label{sec:crystal_velocity}

Crystal growth can be assumed to be limited by the reaction at the crystal-melt interface \citep{Kirkpatrick1975}.
To pass from the melt to the crystal, a molecule must go through an activated state before releasing energy.
The free energy difference between the melt and the activated state is denoted $\Delta G'$ and between the melt and the crystal $\Delta G_{\rm c}$.
We define a  constant rate $\alpha$, equivalent to a diffusion coefficient for transport across the melt-crystal interface, which is defined with the Stokes-Einstein relation as $\alpha = RT_{\rm c} / (3\pi  \mathcal{N}_a  a_0 \mu_{\rm d})$ with $R$ the gas constant and $ \mathcal{N}_a$ the Avogadro number \citep{Kirkpatrick1975}.
Based on the reaction-rate theory, the crystal growth velocity writes
\begin{equation}
    v_{\rm c} =  \frac{RT_{\rm c}}{3\pi  \mathcal{N}_a  a_0^2 \mu_{\rm d}} \left( 1 - \exp{\left(\frac{-\Delta G_{\rm c}}{RT_{\rm c}}\right)}\right).
    \label{eq:growth}
\end{equation}

By assuming that entropy and enthalpy differences are independent of temperature, $\Delta G_{\rm c}$ can be approximated as $M_{\rm d} \mathcal{L} \Delta T_{\rm mc} /T_{\rm m}$ \citep{Wagstaff1968} with $M_{\rm d}$ the molar mass of the liquid.
The small undercooling approximation ($\Delta G_{\rm c} \ll RT_{\rm c}$) is often used to linearized the crystal growth velocity as

\begin{equation}
    v_{\rm c} \simeq \kappa \Delta T_{\rm mc},
    \label{eq:growth-simp}
\end{equation}

\noindent in which we introduce the kinetic undercooling coefficient $\kappa =  M_{\rm d} \mathcal{L} / 3\pi a_0^2 \mu_{\rm d} \mathcal{N}_a T_{\rm m}$.
We note that both expressions of crystal growth do not depend on the problem  geometry and are valid for 2D and 3D growth \citep{Kant2020}.
The mean free path in a liquid is assumed to be of the same order of magnitude as the intermolecular distance. Therefore, $a_0 \sim (M_{\rm d} / \rho_{\rm d} \mathcal{N}_a)^{1/3} \simeq 8 \times 10^{-10}$ m.
The theoretical value of $\kappa$ from equation \eqref{eq:growth-simp} gives $ 1.3 \times 10^{-2}$ m/s/K.

\begin{figure}
    \centering
    \includegraphics[width=\linewidth]{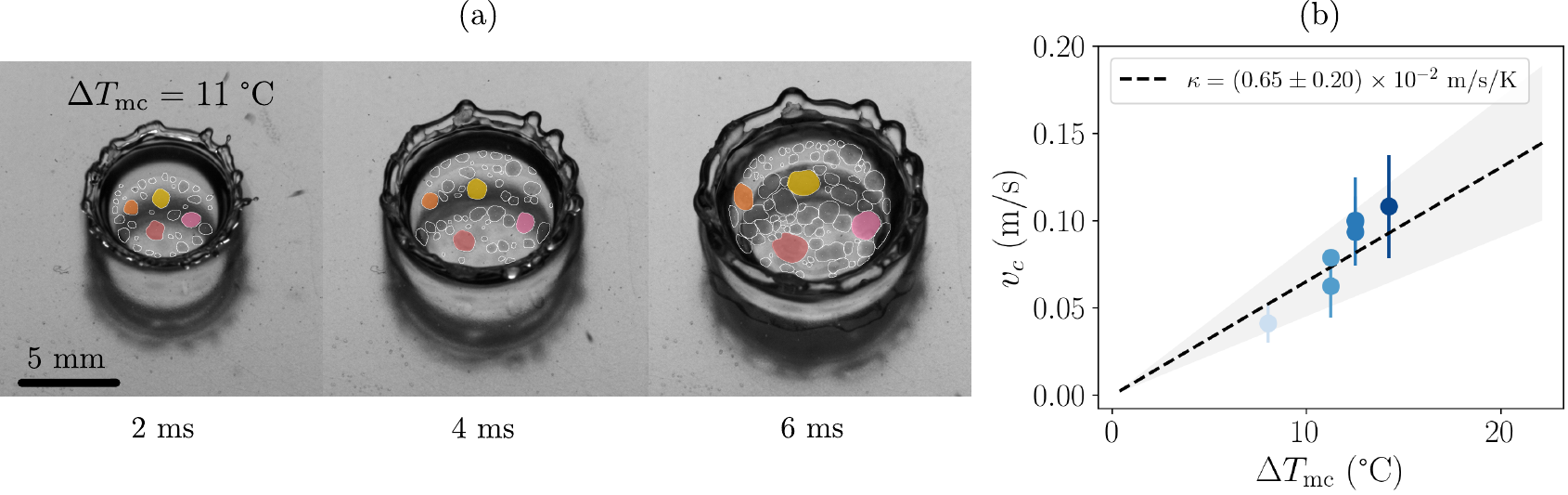}
    \caption{(a) Time series illustrating the evolution and measurements of the crystals for $\Delta T_{\rm mc}=11~^\circ$C.
    The edges of the crystals are indicated by a thin white line and some crystals are coloured for clarity.
    (b) Crystal growth velocity as a function of $\Delta T_{\rm mc}$.
    Each point represents the mean value of individual crystals growing on one experiment, as highlighted on (a).
    The black dashed line is a fit, giving $\kappa = (0.65 \pm 0.20) \times 10^{-2}$ m/s/K.
    The gray area indicates the effect of the uncertainty on $\kappa$.}
    \label{fig:vc}
\end{figure}

Additionally, \citet{Ruiter2017} and \citet{Koldeweij2021} studied crystal growth in hexadecane on solid substrates and they measured experimental values for $\kappa$ of $1.1 \times 10^{-2}$ and $ 0.45 \times 10^{-2}$ m/s/K, respectively.
In a previous study on alkanes solidifying on a liquid substrate \citep{Berry2024}, the value was measured at $0.6 \times 10^{-2}$ m/s/K.
In the present study, the coefficient can be determined experimentally from the images as shown in figure {\ref{fig:vc}}(a) by measuring directly the growth velocity of individual crystals.
In figure {\ref{fig:vc}}(b), we plot these measurements, each point representing a mean value of multiple crystals growing in one experiment, and a linear fit provides $\kappa = (0.65 \pm 0.20) \times 10^{-2}$ m/s/K.

While this small undercooling  approximation operates at the boundary of its conventional validity range, it still provides meaningful insights.
Also, the expression of $v_{\rm c}$ will not be replaced in the rest of the model and the approximation will mainly be used to study the relation between the thermal shock and the number of crystals during the discussion, keeping in mind that the order of magnitude of $\kappa$ is $10^{-2}$ m/s/K and that the crystal growth rate $v_{\rm c}$ is proportional to thermal shock $\Delta T_{\rm mc}$.

\subsection{Number and surface area covered by crystals}

The increase of the number of crystals over a duration ${\rm d}t$ is assumed to be proportional to a nucleation rate $\dot n$, which has the dimension of number of molecules per unit area per time, and to the available surface area $S_{\rm d}(t) - S_{\rm c}(t)$, such as
\begin{equation}
    {\rm d}N_{\rm c}(t) = \dot{n} \left(S_{\rm d}(t) - S_{\rm c}(t)\right)\,{\rm d}t.
    \label{eq:nbr-new}
\end{equation}
The nucleation rate $\dot{n}$ is related to the number of crystals that develop beyond a critical size to become stable and is written as
\begin{equation}
    \dot{n} = A \exp{\left( - \frac{E_{\rm n}}{RT_{\rm c}}\right)}, \label{eq:npoint}
\end{equation}
with $A$ the attempt frequency per unit area and the nucleation activation energy $E_{\rm n} = (16\pi/3) \sigma_{ls}^3f(\theta) \mathcal{N}_a / \Delta G_{\rm c}^2$ \citep{mullin2001,Koldeweij2021}.
We note that $f(\theta)$ is a function of the contact angle between the solid crystal and the substrate, and is accounting for heterogenous nucleation as $E_{\rm het} = f(\theta) E_{\rm hom}$, \textit{i.e} a value of $f(\theta)$ close to 1 in an undercooled liquid is equivalent to homogeneous nucleation whereas a small value of $f(\theta)$ is associated to heterogenous nucleation near the substrate \citep{mullin2001}.
In our approach, we consider heterogeneous nucleation in the drop because the crystals appear on the interface due to the temperature gradient growing as a diffusive process in the drop. The temperature gradient perpendicular to the interface, induced by the cold substrate, dominates over the much more localized in-plane gradient near growing crystals. Consequently, we model nucleation as spatially uniform over the available surface, treating crystals as independent entities. This simplification may lead to an overestimation of the nucleation rate, particularly during the final stages of surface coverage.

We introduce the surface area $\sigma(t,t_0)$ of a single crystal, at a time $t$, that appeared at $t_0\leq t$,
\begin{equation}
    \sigma(t,t_0) = \pi v_{\rm c}^2 (t-t_0)^2,
    \label{eq:surface_area_single_crystal}
\end{equation}
where the crystal growth velocity is given by equation \eqref{eq:growth-simp}.
The surface area $S_{\rm c}(t)$ covered by crystals is the result of the time evolution of each germ, described by equation \eqref{eq:surface_area_single_crystal}, which appeared at a rate ${\rm d}N_{\rm c}(t')$,
\begin{equation}\label{eq:Sc_definition}
    S_{\rm c}(t) = \int^{N_{\rm c}(t)}_{N_{\rm c}(t'=0)} \sigma(t,t') \, {\rm d}N_{\rm c}(t').
\end{equation}
This covered surface area grows until it reaches the surface area of the drop at a time $t_{\rm end}$, \textit{i.e.} $S_{\rm c}(t_{\rm end}) = S_{\rm d}(t_{\rm end})$.

%%%%%%%%%%%%%%%%%%%%%%%%%%%%
\subsection{Dimensionless equations}

To obtain dimensionless equations, we introduce the dimensionless time  $\overline{t} = t/\tau_{\rm m}$, the dimensionless surface $\overline{S_i} = S_i / S_{\rm m}$, where $i=\{{\rm c}, {\rm d}\}$.
Thus, equation \eqref{eq:S_drop} becomes $\overline{S_{\rm d}}(\overline{t}) = 1 - \left( \overline{t} - 1\right)^2$ for $\overline{t}< 1$.
We then obtain the dimensionless form of equation \eqref{eq:nbr-new} on the number of crystals,
 \begin{equation}
    {\rm d}\overline{N_{\rm c}}(\overline{t}) = \left(1 - \left( \overline{t} - 1 \right)^2 - \overline{S_{\rm c}}(\overline{t})\right) {\rm d}\overline{t},
    \label{eq:dimless_dNdt}
\end{equation}
with the dimensionless number of crystals written as
$\overline{N_{\rm c}} = N_{\rm c} / N_{\rm m}$, with $N_{\rm m} = \dot{n} \tau_{\rm m} S_{\rm m}$.

With equation \eqref{eq:Sc_definition} on the surface area covered by crystals, we have
\begin{equation}
     \overline{S_{\rm c}}(\overline{t}) =  \chi^3 \int^{\overline{t}}_{0}  (\overline{t}-\overline{t}')^2  \left( 1 - \left( \overline{t}' - 1 \right)^2 -   \overline{S_{\rm c}}(\overline{t}')\right) {\rm d}\overline{t}',
    \label{eq:dimless_Sc}
\end{equation}
with $\chi^3 = \dot{n} \tau_{\rm m} \cdot \pi v_c^2 \tau_{\rm m}^2 $.
The parameter $\chi^3$ is the nucleation rate multiplied by impact time and by $\pi v_c^2 \tau_{\rm m}^2 $, which is the surface covered by one crystal that grew between $t=0$ and $\tau_{\rm m}$.

To obtain equations \eqref{eq:dimless_dNdt} and \eqref{eq:dimless_Sc} in a form more appropriate for a numerical resolution, we use the Leibniz integral rule, which gives
\begin{subequations}
\begin{align}
    \frac{{\rm d}\overline{S_{\rm c}}}{{\rm d}\overline{t}} &=  2 \chi^3 \int^{\overline{t}}_{0}
      (\overline{t}-\overline{t}')  \left( 1 - \left( \overline{t}' - 1 \right)^2  -  \overline{S_{\rm c}}(\overline{t}')\right)
       {\rm d}\overline{t}',
    \label{eq:dimless-pi_derivative}\\
             \frac{{\rm d}\overline{N_{\rm c}}}{{\rm d}\overline{t}} &= 1 - \left( \overline{t} - 1 \right)^2 - \overline{S_{\rm c}}(\overline{t}) . \label{eq:dimless-xi_derivative}
\end{align}\label{eq:dimless-xi-pi}
\end{subequations}
These equations are complemented by the initial condition $\overline{S_{\rm c}}(0) = \overline{N_{\rm c}}(0) = 0$.
With equation \eqref{eq:S_drop}, the end condition $S_{\rm c}(t_{\rm end}) = S_{\rm d}(t_{\rm end})$ reads
\begin{equation}
    \overline{S_{\rm c}}(\overline{t}_{\rm end}) = 1 - \left( \overline{t}_{\rm end} - 1\right)^2 \qquad{\rm with}\quad \overline{t}_{\rm end}< 1.
    \label{eq:condition_end}
\end{equation}

\noindent If  $\overline{t}_{\rm end} > 1$, crystals have not covered the drop during the opening phase of the impact and the model ceased to be valid due to the description adopted for the hydrodynamics.

In the next Section, we solve these equations, discuss the results of the model, and compare the predictions with the experimental measurements.

\begin{figure}
    \centering
    \includegraphics[width=1\linewidth]{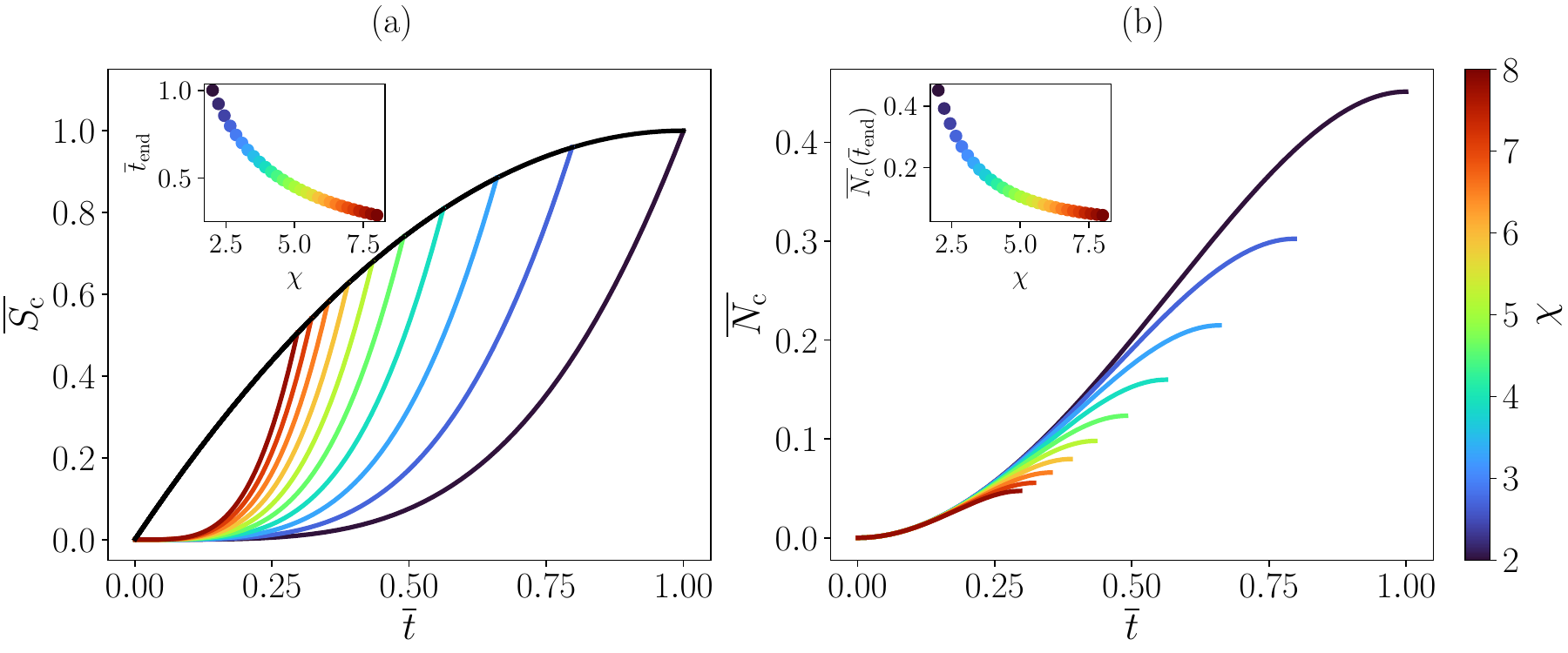}\\
    \caption{Variation of the solutions of equations~\eqref{eq:dimless-xi-pi} with $\chi$.
(a) Dynamics of the surface area covered by the crystals.
The black line represents the dimensionless evolution of the contact area area.
(b) Time evolution of the number of crystals.
}
    \label{fig:solution}
\end{figure}

%\newpage\clearpage
%%%%%%%%%%%%%%%%%%%%%%%%%%%%
%
%%%%%%%%%%%%%%%%%%%%%%%%%%%%
\section{Discussion}\label{sec:discussion}

%%%%%%%%%%%%%%%%%%%%%%%%%%%%
\subsection{Numerical solution and effect of the physical parameters}

Now, we solve numerically the differential equations \eqref{eq:dimless-xi-pi} with the library scipy in Python until the condition \eqref{eq:condition_end} is satisfied.
In figure \ref{fig:solution}, we plot the dimensionless surface covered by crystals and the dimensionless number of crystals as a function of time.

Both $S_{\rm c}$ and $N_{\rm c}$ increase, which is expected as the only mechanism taken into account is the apparition of crystals.
The black line represents the surface area of the drop.
The surface covered increases slowly in the beginning, as there are only a few crystals at first, but with more and more crystals appearing, the surface covered increases faster until it reaches the contact area.
The curve for the number of crystals makes an S-shape.
At first, the production rate of crystals increases as more surface becomes available with the impact deformation, then the growing crystals start to cover a significant part of the surface, which limits the production rate.
The inflection point corresponds to the maximum of $S_{\rm d}(t) - S_{\rm c}(t)$ in  equation \eqref{eq:nbr-new}.

We note that the dimensionless number of crystals is defined as $\overline{N_{\rm c}} = N_{\rm c} / N_{\rm m}$, with $N_{\rm m} = \dot{n} \tau_{\rm m} S_{\rm m}$ (equation \eqref{eq:dimless_dNdt}).
The number $N_{\rm m}$ is the number of crystals that could be created at a nucleation rate $\dot{n}$ during the impact time $\tau_{\rm m}$ if the entire maximum surface $S_{\rm m}$ was available during that time.
It is an overestimation of the number of crystals which explains why the final dimensionless  number of crystals $\overline{N_{\rm c}}(\overline{t}_{\rm end})$, in the inset of figure \ref{fig:solution}(b), does not reach 1.

The parameter $\chi$ is introduced in equation \eqref{eq:dimless_Sc} as $\chi^3 = \dot{n}\tau_{\rm m} \cdot \pi v_{\rm c}^2 \tau_{\rm m}^2$.
Considering the expression of $\tau_{\rm m}$ \eqref{eq:tmax}, $\chi$ varies linearly with $v_0$.
The growth rate $v_{\rm c}$ is assumed to be linear with $\Delta T_{\rm mc}$ \eqref{eq:growth-simp} and the nucleation rate $\dot{n}$ also increases with the thermal shock \eqref{eq:npoint}.
A larger thermal shock leads to faster crystal growth, meaning that the area covered by crystals reaches the drop area much faster and fewer crystals have time and space to appear.
This is consistent with the trend observed in experiments, where large thermal shock leads to fewer crystals.

Qualitatively, this model describes well the behaviour observed experimentally and presented in figure \ref{fig:observations}.
In the next section, we compare more precisely the data and the model.

%%%%%%%%%%%%%%%%%%%%%%%%%%%%
\subsection{Comparison with the experiments}
 \begin{figure}
    \centering
   \includegraphics[width=1\linewidth]{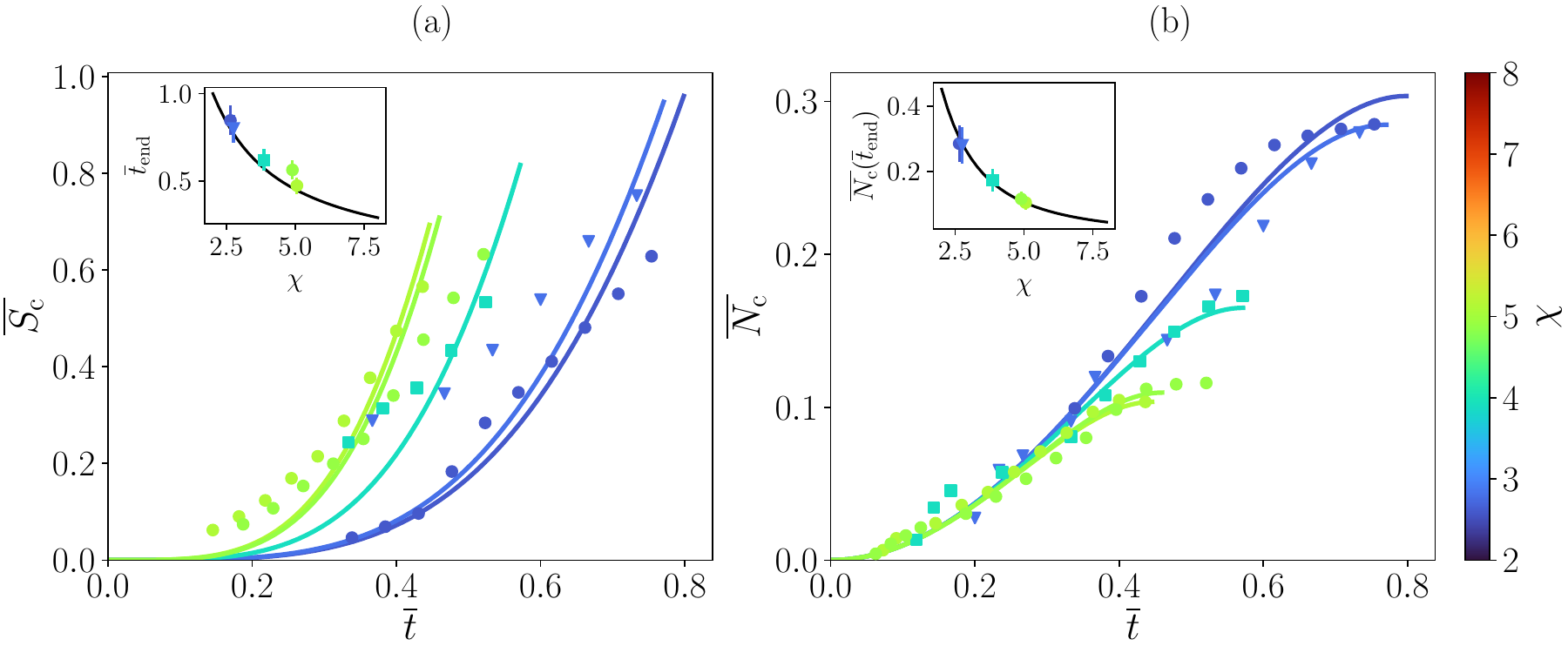}
    \caption{Dimensionless curves for (a) the surface area and (b) the number of crystals with experimental measurements as symbols and numerical solutions as lines.
    The inset in (a) shows the dimensionless time to get a full coverage as a function of $\chi$ and the inset in (b) is for the final number of crystals.
    Black curves represent the model.
    The markers indicate  impact velocity of $v_0 = 2.8$ m/s (circle), $v_0 = 2.0$ m/s (square), and $v_0 = 1.4$ m/s (downward triangle).
    }
    \label{fig:results}
\end{figure}

We are now in a position to compare the predictions of the model with the experimental results.
The different parameters in the model are $S_{\rm m}$, $\tau_{\rm m}$, and $N_{\rm m}$.
The first two parameters are solely linked to hydrodynamics.
Their values are taken directly from each experiment to compute each data point in a dimensionless form in figure~{\ref{fig:results}}.
The third parameter $N_{\rm m}$ depends on the kinetic coefficient $\kappa$ and the nucleation rate $\dot{n}$.

We adjust these two parameters to minimize the difference between experiments and numerical solution for the final number of crystals $\overline{N_{\rm c}} (\overline{t}_{\rm end})$.
Even if we do not measure the number of crystals right to the end, $\overline{N_{\rm c}}$ plateaus (figure \ref{fig:results}(b)), which means that the measured value is close to the final number of crystals.
The dimensionless dynamics of both the model and the data are given in figure \ref{fig:results}.
The quality of the fit is given in the inset of figure \ref{fig:results}(b), as a function of $\chi$.
This inset also shows how the final number of crystals varies with $\chi$, therefore with thermal shock and impact velocity.
We first discuss how the experimental dynamics compare to the model, and we then address the values obtained through the fit.

The dynamics of both $\overline{S_{\rm c}}$ and $\overline{N_{\rm c}}$ are captured reasonably well by the model.
Regarding the covered surface, the model underestimates the first measurements of the bright green points, which correspond to the largest thermal shock.
On the corresponding videos, we observe generally one or two crystals appearing almost immediately near the center of the drop, which might have been created due to a defect or a singularity point when the contact between the drop and the substrate is made and is therefore not captured by the model.
The number of crystals seems to be better described by the model than the covered surface, with the S-shape behaviour reasonably well captured.

The nucleation rate $\dot{n}$ is adjusted with $f(\theta)$ in the energy activation and $A$ the exponential pre-factor. The value of $f(\theta) = 0.01$ seems to indicate that the nucleation is on the interface rather than in the bulk, which agrees with our assumptions \cite{mullin2001}.
The pre-factor $A = 3 \times 10^9$ m$^{-2}$.s$^{-1}$ is reasonable given the typical rate of crystal appearance.
In our range of thermal shock, the exponential value in the nucleation rate expression is in [0.65-0.85] and increases almost linearly with the thermal shock, which suggests $\chi^3 \sim \dot{n}v_{\rm c}^2 \tau_{\rm m}^3 \propto \Delta T_{\rm mc}^3$.

The value of $\kappa$ obtained through the fit, $1.2 \times 10^{-2}$ m/s/K, is of the same order of magnitude as the values previously obtained in the literature, which means that our model is consistent.
However, this value is different by a factor 2 from the value measured through direct measurements (figure {\ref{fig:vc}}).
The assumption of the absence of crystal overlapping or mutual interaction in the model is not explaining this difference.
A possible explanation for this difference would be that the equations for $v_{\rm c}$ and $\dot{n}$ are not perfectly describing the system. Since the solution comes from an integro-differential equation, errors tend to accumulate.
This explanation would be supported by the disagreement observed on $\overline{S_{\rm c}}$ at short times.
However, the lack of measurements at that moment prevents us to precisely identify the origin and to refine the model.

Nevertheless, the temperature dependence of the number of crystals is confirmed both qualitatively and quantitatively with the model, which is already satisfactory.

% \newpage\clearpage
%%%%%%%%%%%%%%%%%%%%%%%%%%%%
% CCL
%%%%%%%%%%%%%%%%%%%%%%%%%%%%
\section{Conclusion}

We investigated the nucleation of crystals at an interface, more specifically in the case of an alkane drop impacting a cold thin liquid substrate.
Through high-speed imagery and direct measurements, we observed the crystals and their evolution, noticing a dependence with the thermal shock $\Delta T_{\rm mc}$ defined with the contact temperature.
We developed a model based on crystal growth appearing on the available surface area.
This area evolves in time by the competing effects of drop spreading upon impact and the surface coverage of crystals growing at a velocity $v_{\rm c} = \kappa \Delta T_{\rm mc}$.
The numerical solution captures well the observed behaviour over time for both the covered area and the number of crystals.
Furthermore, we were able to quantitatively compare the model with measurements with two fitting parameters $\kappa$ and $\dot{n}$.
We note that in this study $\kappa$, the kinetic coefficient of crystal growth rate, is comparable to previous studies.
We can thus successfully describe crystal formation on liquid interfaces following a drop impact and we can explain the variation of the number of crystals with the thermal shock.

These results rationalize the formation of the solid layer between the drop and the bath, which is responsible for the final morphology of the drop \citep{Berry2024}.
In future work, it will be interesting to consider more complex systems, such as the crystallization of surfactant by salt where the two liquids are miscible, prone to Marangoni effects, and with a solidification process driven by a solubility threshold \citep{Kharlamova2024}.

%%%%%%%%%%%%%%%%%%%%%%%%%%%%
% END SECTIONS
%%%%%%%%%%%%%%%%%%%%%%%%%%%%

\backsection[Supplementary data]{\label{SupMat}Supplementary material and movies are available at \\https://doi.org/10.1017/jfm.2019...}

\backsection[Acknowledgements]{We thank V. Thi\'evenaz for suggesting the apparatus for the film thickness measurement.}

\backsection[Funding]{This work received a financial support by  Investissements d’Avenir, LabEx PALM (ANR-10-LABX-0039-PALM).}

\backsection[Declaration of interests]{The authors report no conflict of interest.}

%\backsection[Data availability statement]{The data that support the findings of this study are openly available in [repository name] at http://doi.org/[doi], reference number [reference number]. See JFM's \href{https://www.cambridge.org/core/journals/journal-of-fluid-mechanics/information/journal-policies/research-transparency}{research transparency policy} for more information}

\backsection[Author ORCIDs]{F. Boulogne https://orcid.org/0000-0003-2617-4554, A. Salonen https://orcid.org/0000-0002-8286-3250}

\backsection[Author contributions]{Authors may include details of the contributions made by each author to the manuscript'}

% \newpage
%\clearpage
%%%%%%%%%%%%%%%%%%%%%%%%%%%%
% APPENDIX
%%%%%%%%%%%%%%%%%%%%%%%%%%%%
\appendix

\section{Properties}\label{appA}
\clearpage
%%%%%%%%%%%%%
\begin{table}
\centering
    \begin{tabular}{lcccccccc}
    \hline
    & $T_{\rm m}$ & \multicolumn{2}{c}{Density} & Surface tension & Dynamic viscosity &   \\
     &  ($^{\circ}$C) & \multicolumn{2}{c}{$\left(\mathrm{kg} / \mathrm{m}^3\right)$}  &  (mN/m) &   (mPa.s) &  \\
    \hline
     & & Liquid  & Solid  &  &  & \\
     Hexadecane  & 18.1 & {743} & 886 & {27.2} & {3.8}  & \\
     NaCl 23.3\%wt & -21 &  {1184} & &  {85} &  {2.6} &\\
    \hline
    \end{tabular}
    %
    %
    % TABLE #2
    %
    %

    \begin{tabular}{lccccccccc}

    \hline
      &  Latent heat  &\multicolumn{2}{c}{Specific heat capacity} & \multicolumn{2}{c}{Conductivity} & \multicolumn{2}{c}{Diffusivity} & \multicolumn{2}{c}{Effusivity}\\

     &  $ (\mathrm{kJ} / \mathrm{kg})$&\multicolumn{2}{c}{$(\mathrm{kJ} / \mathrm{kg}-\mathrm{K})$}  &  \multicolumn{2}{c}{$(\mathrm{W} / \mathrm{m}-\mathrm{K})$} &
                \multicolumn{2}{c}{$\times 10^{-7}\left(\mathrm{m}^2 / \mathrm{s}\right)$} & \multicolumn{2}{c}{$(\mathrm{W} \sqrt{s} / \mathrm{m^2}-\mathrm{K})$}\\
    \hline
     Hexadecane &  236 & 2.20 & 1.81 & 0.14 & 0.22 & {0.86} & {1.37}  & {478}& {594}\\
     NaCl 23.3\%wt &  & 3.3 & & 0.55 &  & {1.41} &  & {1466} &\\
    \hline
    \end{tabular}
    \caption{Physical properties used in the present study for hexadecane  taken from \citet{Yaws1999,Kulkarni2024,Berry2024},  and NaCl 23.3~\%wt. brine taken from \citet{Chen1982,Abdulagatov1994,Aleksandrov2013,Ramalingam2012}.
    The alkane properties are taken at $T_{\rm d}$ = 20 $^{\circ}$C, except the solid properties taken at their respective melting temperature $T_{\rm m}$.
    The NaCl brine properties are taken at a mean temperature of about 0~$^{\circ}$C.
    }
    \label{t:prop}
\end{table}

\newpage
%%%%%%%%%%%%%%%%%%%%%%%%%%%%
% BIBLIOGRAPHY
%%%%%%%%%%%%%%%%%%%%%%%%%%%%
\bibliographystyle{jfm}
\bibliography{jfm}

\end{document}